\documentclass[reprint,superscriptaddress,showpacs,nofootinbib,amsmath,amssymb,aps,prl]{revtex4-1}
\usepackage{graphicx}% Include figure files
\usepackage{dcolumn}% Align table columns on decimal point
\usepackage{bm}% bold math
\usepackage[utf8]{inputenc}
\usepackage[frenchb]{babel}
\usepackage{braket}
\usepackage[T1]{fontenc}

\begin{document}

\preprint{APS/123-QED}

\title{Spin-orbit coupled depairing of a dipolar biexciton superfluid}

\author{S. V. Andreev}
\email[Electronic adress: ]{Serguey.Andreev@gmail.com}
\affiliation{National Research Center "Kurchatov Institute" B.P.\ Konstantinov Petersburg Nuclear Physics Institute, Gatchina 188300, Russia}

\date{\today}

\begin{abstract}  
We consider quantum phase transitions in a system of bright dipolar excitons which can form bound pairs (dipolar biexcitons). We assume a narrow resonance in the interaction of excitons with opposite spins. At sufficiently large density a resonant exciton superfluid transforms into a superfluid of biexcitons. The transition may be either of the first or the second kind. The average relative momenta of excitons in the pairs being beyond the light cone, the transition should be accompanied by reduction of the photoluminescence intensity. Effective magnetic fields due to the long-range exchange splitting of non-radiative exciton states induce broadening of the biexciton resonance. The fields shift the position of the gap in the elementary excitation spectrum to a circle of degenerate minima in the $\bm{k}$-space. Closing the new gap defines a second order phase transition into a mixture of counter-propagating plane-wave excitonic condensates polarized linearly in the direction perpendicular to their wavevectors. In the resonance energy vs density phase diagram the novel phase intervenes between the dark biexciton and radiative exciton superfluids. We conclude that formation of a BCS-like biexciton condensate induces correlated alignment of the effective magnetic fields and excitonic spins. We outline important differences of the emergent mechanism from the phenomenon of spin-orbit (SO) coupled Bose-Einstein condensation. We expect existence of analogous mechanisms in SO-coupled fermionic superfluids and superconductors. 
\end{abstract}


\maketitle

\textit{Introduction.} Excitonic molecules (biexcitons) have attracted great interest since the very beginning of the quest for the Bose-Einstein condensation (BEC) in solids \cite{Hanamura}. The concept of a biexciton is an extension of the analogy between an exciton and a hydrogen atom to the case of two quasiparticles: exchange of the constituent fermions in the spin-singlet configuration stabilizes a bound state of bosons \cite{Moskalenko, Lampert, Nikitine, Agranovich, Brinkman1, Kulakovskii}. At low temperatures a biexciton condensate should compete with a condensate of unpaired excitons \cite{Noziers}. 

\begin{figure}
\includegraphics[width=0.9\linewidth]{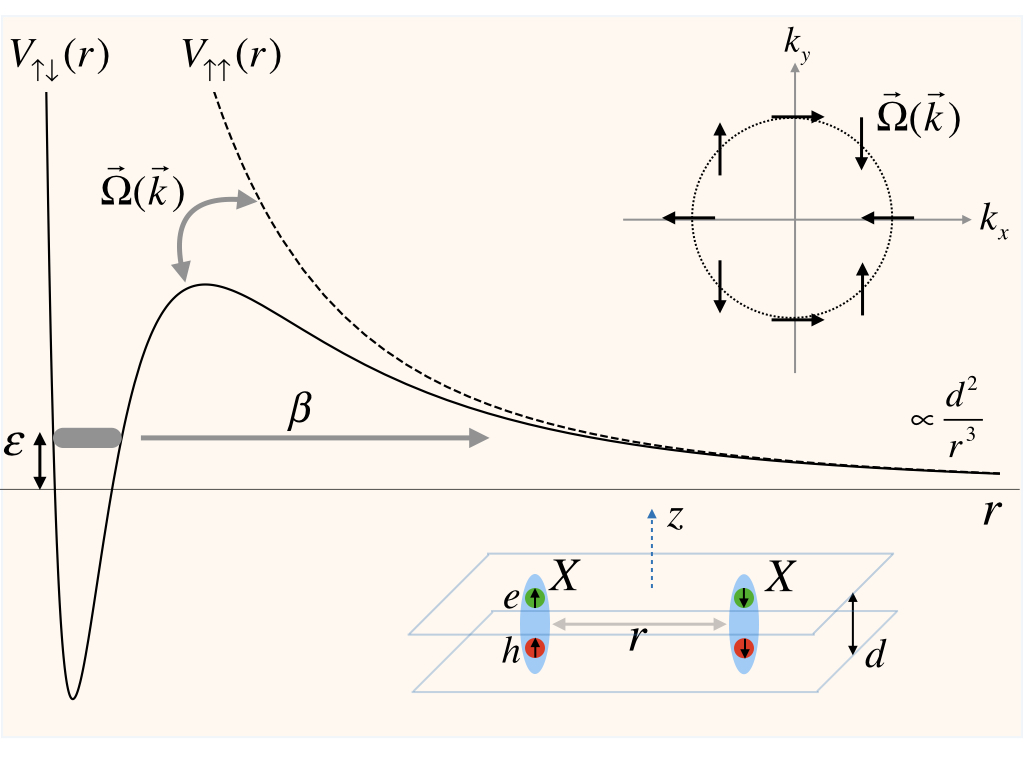} 
\caption{\label{Fig0}Sketch of the two-body interaction potentials for excitons with parallel (dashed line) and anti-parallel (solid line) spins. The latter features a biexciton resonance with energy $\varepsilon$ and natural width $\beta$ defined by tunneling of excitons under the barrier due to the dipolar repulsion. The inset in the upper right corner shows orientations of the effective magnetic field $\bm\Omega(\bm k)$ with respect to the exciton wavevector $\bm k$. The field $\bm\Omega(\bm k)$ couples the $\ket{\uparrow\downarrow}$ and  $\ket{\uparrow\uparrow}$ (or $\ket{\downarrow\downarrow}$) scattering channels. The bottom inset shows a pair of dipolar excitons (X) formed of spatially separated electrons (e) and holes (h).}
\end{figure}   

With the advent of semiconductor technology, well-resolved biexciton lines have been observed in photoluminescence (PL) of quantum wells (QW's) \cite{Miller,Wolfe, Spiegel} and, more recently, atomically thin layers of transition metal dichalcogenides (TMD's) \cite{XXinTMD1,XXinTMD2,XXinTMD3}. Bilayer structures or wide QW's subjected to a transverse electric field host dipolar excitons with long radiative lifetimes \cite{High2012, Dubin, BarJoseph, Fogler2014, Rivera2015}. The inset in Fig. \ref{Fig0} shows schematically a pair of dipolar excitons each composed of an electron and a hole residing in a couple of layers spatially separated in the transverse direction (the $z$-axis). A dipolar exciton possesses a permanent dipole moment oriented along $z$. Besides, it also has a pseudospin defined by $z$-projections of the fermionic spins and valley indices. Among a great variety of excitonic species only the so-called bright excitons characterized by two possible pseudospin states ($\ket{\uparrow}$ or $\ket{\downarrow}$) may recombine emitting a photon (right- or left-circularly polarized, respectively) \cite{Colloquium}. The typical biexcitonic PL line corresponds to a bound state of two bright excitons with opposite spins (\textit{i.e.}, $\ket{\uparrow}$ and $\ket{\downarrow}$). Many-body quantum phases and phase transitions in a system of bright dipolar excitons which can form biexcitonic molecules constitute the subject of this Letter. Below we discuss important specifics of the problem and introduce our model.

For large separation between the layers dipolar repulsion prevents formation of biexcitons and favours excitonic superfluidity \cite{Zimmerman, Fogler}. Before completely disappear, however, a dipolar biexciton has been predicted to transform into a resonance \cite{Andreev2015, RP}. As opposed to a true bound state, the resonance may have a positive energy $\varepsilon$ and a finite width $\beta$ defined by tunnelling of excitons under the dipolar potential barrier. As one increases the density, a resonantly paired exciton superfluid (X) undergoes a quantum phase transition to a superfluid of biexcitons (XX). The dipolar biexcitons are stabilized in the media by mean-field repulsion, single excitons occurring only as gapped elementary excitations of the superfluid.

Extended spatial coherence, which may be accessed by shift-interferometry measurements of the PL \cite{High2012, Dubin}, is suppressed in the XX phase due to relative motion of excitons within a molecule \cite{Andreev2015}. For the same reasons, one may expect suppression of the PL intensity: as the effective binding tightens, the excitons are extruded outside of the "light cone". The latter defines a circle of a radius $q_0$ in 2D momentum space, inside which a bright exciton can recombine emitting a photon \cite{Colloquium}. An emergent property of a non-radiative exciton is the fine structure due to the long-range exchange interaction between the electron and the hole \cite{Gupalov, Glazov}. The single-exciton Hamiltonian accounting for the long-range exchange has the form 
\begin{equation}
\label{H0}
\begin{split}
\hat H_0&=\sum_{\bm p, \sigma=\uparrow,\downarrow}\left(\frac{\hbar^2 p^2}{2m}+ \frac{\hbar \upsilon p}{2}\right)\hat a_{\sigma, \bm p}^\dagger \hat a_{\sigma, \bm p}\\
&+\sum_{p, \varphi}\frac{\hbar \upsilon p}{2}(e^{-2i\varphi}\hat a_{\uparrow,\bm p}^\dagger \hat a_{\downarrow,\bm p}+\mathrm{H.c.}),
\end{split}
\end{equation}
where $\bm p=(p,\varphi)$ is the 2D wavevector of the exciton translational motion and one assumes $p\gg q_0$. The lower-energy eigenstate 
$\psi_{\bm p}^\perp(\bm r)=e^{i\bm p\bm r}(-e^{-2i\varphi}\ket{\uparrow}+\ket{\downarrow})/\sqrt{2S}$
of \eqref{H0} is a  plane wave polarized linearly in the direction perpendicular to $\bm p$ in the quantization area $S$. Its dispersion $E_\perp(p)=\hbar^2 p^2/2m$ is split from the upper branch $E_\parallel(p)=\hbar^2 p^2/2m+\Delta_\mathrm{LT}(p)$ polarized along $\bm p$ by the amount $\Delta_\mathrm{LT}(p)=\hbar \upsilon p$ known as the longitudinal-transverse (LT) splitting \cite{Sham, Gupalov, Glazov}. The LT splitting parameter is given by
\begin{equation}
\label{upsilon} 
\upsilon=1/\tau q_0,
\end{equation}
where $\tau$ is the radiative lifetime of excitons. Note, that in contrast to the single-particle dispersions in the presence of conventional spin-orbit (SO) coupling \cite{Dresselhaus, BychkovRashba, DyakonovKachorovskii, Galitski, Stringari, Ketterle, Hanle, Spielman}, the function $E_\perp(p)$ is \textit{monotonous}.

\begin{figure}
\includegraphics[width=0.7\linewidth]{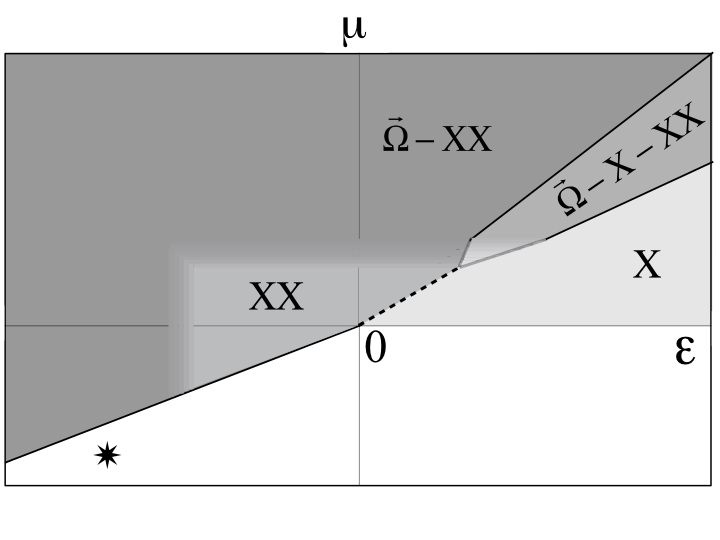}  
\caption{\label{Fig1}Sketch of the phase diagram of a resonantly paired exciton condensate in the narrow-resonance limit ($\varepsilon\gg \beta$). For sufficiently large $\lvert \varepsilon\rvert$ the biexciton phase (XX, gray and $\bm \Omega $-XX, dark gray colors, respectively) is characterized by suppressed PL intensity as compared to the exciton condensate (X, light gray). The exchange fields acting on the non-radiative excitons split the first-order X/XX phase boundary (dashed line) into two second-order (solid) lines representing the boundaries of a new $\bm \Omega $-X-XX phase. The star indicates position of a typical experiment with non-dipolar excitons (see, e.g., Ref. \cite{Wolfe}).}
\end{figure}

The transverse polarization of $\psi_{\bm p}^\perp$ corresponds to alignment of the exciton pseudospin along the effective magnetic field $\bm\Omega(\bm p)=(\Omega_x, \Omega_y)$, with $\Omega_x(\bm p)=-\upsilon p \cos(2\varphi)$ and $\Omega_y(\bm p)=-\upsilon p \sin(2\varphi)$ (Fig. \ref{Fig0}). In this Letter, we show that in a dark biexciton superfluid (which we denote $\bm\Omega$-XX) the microscopic fields $\bm\Omega(\bm p)$ unify to produce a new efficient pair-breaking mechanism. The fields shift the position of the gap in the $\bm{\Omega}$-XX elementary excitation spectrum from $\bm p=0$ to a circle of roton-like minima at $\lvert \bm p\rvert=p_{\bm \Omega}$, as shown in Fig. \ref{Fig1}. This spectacular $N$-body effect persists in a properly defined thermodynamic limit $N\rightarrow\infty$, $S\rightarrow\infty$ and $\tau\rightarrow\infty$, where the LT splitting of single-particle states \eqref{upsilon} is vanishingly small. The magnitude of $p_{\bm \Omega}$ [Eq. \eqref{RotonPosition}] is defined by the balance between the kinetic energy and interaction of the exciton spin with the exchange field enhanced by the molecular coherence. As a result, depairing of the biexciton superfluid occurs into a superposition of $\bm p_{\bm \Omega}$ and $-\bm p_{\bm \Omega}$ plane-wave excitonic condensates polarized transversely to $\bm p_{\bm \Omega}$ [Eq. \eqref{PsiX}]. We denote this novel phase as $\bm{\Omega}$-X-XX. Linearly polarized excitonic waves in the $\bm{\Omega}$-X-XX phase originate from resonant depairing of a BCS-like condensate of loosely bound excitonic molecules. The wavevector $p_{\bm \Omega}$ [Eq. \eqref{PlaneWaveVector}] has its maximum value at the $\bm{\Omega}$-XX/$\bm{\Omega}$-X-XX phase transition boundary and approaches the light cone $q_0$ as one decreases the density. At that point the $\bm{\Omega}$-X-XX phase turns into the radiative exciton superfluid X.

\textit{The model.} The full many-body Hamiltonian reads $\hat H=\hat H_0+\hat V$, where                  
\begin{equation}
\label{V}
\hat V=\frac{1}{2S}\sum_{\textbf p_1,\textbf p_2,\textbf{q},\sigma,\sigma^\prime}\hat a_{\sigma, \textbf p_1+\textbf q}^{\dag} \hat a_{\sigma^\prime,\textbf p_2-\textbf q}^{\dag} V_{\sigma\sigma^\prime}(\textbf{q})\hat a_{\sigma, \textbf p_1}\hat a_{\sigma^\prime,\textbf p_2}
\end{equation}
is the two-body interaction in a binary mixture of interacting bosons and $V_{\sigma\sigma'}(\bm q)=\int e^{-i\bm q\bm r}V_{\sigma\sigma'}(\bm r)d\bm r$. The potentials $V_{\uparrow\uparrow}(\bm r)$ and $V_{\downarrow\downarrow}(\bm r)$ are repulsive at all distances, whereas $V_{\uparrow\downarrow}(\bm r)$ features a resonance, characterized by its energy $\varepsilon$ and width $\beta$ (Fig. \ref{Fig0}). Throughout this article we assume $\varepsilon\gg\beta$.

For a typical MoS$_2$-based heterostructure the momentum cutoff imposed by the light cone corresponds to the pair energy on the order of $\varepsilon\sim10$ $\mu$eV. At the X-XX transition one has $\varepsilon\sim \mu$. The chemical potential of a 2D exciton condensate can be estimated by using the transcendental formula $\mu=4\pi/\ln(E_a/2\mu) \hbar^2 n/m$ \cite{Utesov}, where $m$ is the exciton mass, $E_a=\hbar^2/ma^2$ with $a$ being on the order of the exciton Bohr radius $a_\mathrm{X}$. We thus obtain $n\sim 10^{10}$ cm$^{-2}$ for the reference exciton density separating the radiative and dark regimes. At such densities the system shares analogies with gaseous atomic superfluids \cite{Safonov, Dalibard}.

\textit{Radiative regime.} It is instructive to discuss first the infinitely narrow resonance limit ($\beta=0$) of the radiative X and XX phases. The LT splitting is absent and we may describe the ground state by the following system of Gross-Pitaevskii equations
\begin{equation}
\label{GP}
\begin{split}
2\mu \Psi_{\mathrm B}&=(\varepsilon+g_{\mathrm B} \lvert \Psi_{\mathrm B}\rvert^2+g_{\uparrow \mathrm B}\lvert \Psi_{\uparrow}\rvert^2+g_{\downarrow \mathrm B}\lvert \Psi_{\downarrow}\rvert^2)\Psi_{\mathrm B}\\
\mu \Psi_\sigma&=(g_{\sigma\uparrow}\lvert \Psi_{\uparrow}\rvert^2+g_{\sigma\downarrow}\lvert \Psi_{\downarrow}\rvert^2+g_{\sigma\mathrm{B}}\lvert \Psi_{\mathrm B}\rvert^2) \Psi_\sigma, 
\end{split}
\end{equation}
where  $\Psi_{\sigma}$ and $\Psi_{\mathrm B}$ stand for the exciton and biexciton order parameters, respectively, and $\sigma=\{\uparrow,\downarrow\}$. One has $2\lvert \Psi_{\mathrm B}\rvert^2+\lvert \Psi_{\uparrow} \rvert^2+\lvert \Psi_{\downarrow} \rvert^2=n$. The effective interactions $g_{\sigma\sigma^\prime}$ account both for the short-range part and the dipolar tail of the bare potentials $V_{\sigma\sigma^\prime}(\bm r)$ \cite{Utesov}. The phenomenologically introduced potentials $g_{\sigma\mathrm B}$ and $g_{\mathrm B}$ are positive constants on the order of $g_{\sigma\sigma^\prime}$. By symmetry, we assume $g_{\uparrow\uparrow}=g_{\downarrow\downarrow}\equiv g_{\mathrm X}$, $g_{\uparrow\mathrm{B}}=g_{\downarrow \mathrm{B}}\equiv g_{\mathrm X \mathrm{B}}$ and $\lvert \Psi_{\uparrow} \rvert=\lvert \Psi_{\downarrow}\rvert$. Miscibility of excitons with different spins requires $g_{\uparrow\downarrow}<g_{\mathrm X}$. We may let $g_{\uparrow\downarrow}\equiv0$ without loss of generality. 

Solving the system \eqref{GP} yields the phase diagram shown in Fig. \ref{Fig1}. The region $\varepsilon<0$ corresponds to a true bound state, as is realized, e.g., with non-dipolar excitons. Condensation here occurs directly into the XX phase with the condensate density growing as $\lvert \Psi_{\mathrm B}\rvert^2=(2\mu-\varepsilon)/g_\mathrm{B}$. Signatures of such second-order transition in the exciton thermodynamics were reported in Ref. \cite{Wolfe}.

As the distance $d$ between the electron-hole layers crosses the threshold value $d_c\sim a_{\mathrm X}$, the bound state turns into a resonance. Close to $d_c$ one has $\varepsilon\propto d-d_c$ \cite{RP}. In the region $\varepsilon>0$ the condensation starts in the X phase at the second-order phase transition line $\mu=0$. Depending on whether $\sqrt{g_{\mathrm X} g_{\mathrm B}}<g_{\mathrm X \mathrm{B}}$ or $\sqrt{g_{\mathrm X} g_{\mathrm B}}>g_{\mathrm X \mathrm{B}}$, the subsequent transition to the XX phase may be either of the first or the second kind, respectively. The second-order transition occurs via an intermediate mixed phase (X-XX), where a fraction of biexcitons is dissolved in the X superfluid. The X/X-XX second-order boundary is located at $\mu=\varepsilon/(2-g_{\mathrm X \mathrm B}/g_{\mathrm X})$. As one increases the chemical potential $\mu$, the biexciton fraction grows and at 
\begin{equation}
\label{XX/X-XX}
\mu=\varepsilon/(2-g_{\mathrm B}/g_{\mathrm X\mathrm B})
\end{equation} 
the X-XX mixture continuously turns into the XX superfluid. Assuming that the effective interactions are inversely proportional to the reduced mass of the particle relative motion \cite{Utesov}, one may conclude that the transition should be first order (dashed line in Fig. \ref{Fig1}). The situation changes in the dark regime, as we show below.

\textit{SO coupled depairing of a dark biexciton superfluid.} We now consider $\varepsilon>0$ sufficiently large for the XX phase to be dark. In this case the excitons are subjected to the exchange fields [Eq. \eqref{H0}]. We notice that at the two-body level the exchange field $\bm\Omega(\bm p)$ couples the $\ket{\uparrow\uparrow}$ and $\ket{\downarrow\downarrow}$ repulsive scattering channels to the $\ket{\uparrow\downarrow}$ pairing channel. One may also say that the exchange destroys a biexciton by flipping the spin of either of the two excitons. This motivates us to consider the following BCS-like Hamiltonian [formal derivation starting from Eq. \eqref{H0} can be found in Methods]
\begin{widetext}
\begin{equation}
\label{OmegaHamiltonian}
\begin{split}
\hat H_{\bm\Omega}&=E_\mathrm{B}+\sum_{\bm p, \sigma=\uparrow,\downarrow}\xi_{\bm p}\hat a_{\sigma, \bm p}^\dagger \hat a_{\sigma, \bm p}+\frac{g_{\mathrm X}}{2S}\sum_{\textbf p_1,\textbf p_2,\textbf{q},\sigma}\hat a_{\sigma, \textbf p_1+\textbf q}^{\dag} \hat a_{\sigma,\textbf p_2-\textbf q}^{\dag} \hat a_{\sigma, \textbf p_1}\hat a_{\sigma,\textbf p_2}\\
&+\sqrt{N_{\mathrm B}}\hbar\upsilon\sum_{p,\varphi}[\phi_{\bm p}p(e^{-2i\varphi}\hat a_{\uparrow, \bm p}^\dagger \hat a_{\uparrow, -\bm p}^\dagger+e^{2i\varphi}\hat a_{\downarrow, \bm p}^\dagger \hat a_{\downarrow, -\bm p}^\dagger)+\mathrm{H.c.}]-\sqrt{N_{\mathrm B}}\alpha\sum_{\bm p}(\hat a_{\uparrow, \bm p}^\dagger \hat a_{\downarrow, -\bm p}^\dagger+\mathrm{H.c.}).
\end{split} 
\end{equation}
\end{widetext}
Here
\begin{equation}
E_\mathrm{B}=(\varepsilon-2\mu)N_{\mathrm B}+\frac{g_{\mathrm B}N_{\mathrm B}^2}{2S}
\end{equation}
is the (grand canonical) energy of the molecular condensate and
\begin{equation}
\label{xi}
\xi_{\bm p}=\frac{\hbar^2 p^2}{2m}+ \frac{\hbar \upsilon p}{2}+g_{\mathrm X \mathrm{B}}\lvert \Psi_{\mathrm B}\rvert^2-\mu
\end{equation}
is the energy of unpaired excitons in the mean-field potential produced by the molecules. The occupation number $N_{\mathrm B}$ is related to the molecular order parameter by $\lvert \Psi_{\mathrm B}\rvert^2=N_{\mathrm B}/S$. The last term with 
\begin{equation}
\alpha= \sqrt{\frac{\hbar^2\beta}{2\pi m S}}
\end{equation}
accounts for the finite width $\beta$ of the resonance due to dissociation through the dipolar potential barrier \cite{RP}.

One may see, that the presence of a molecular condensate enhances the exchange field by the factor $\sqrt{N_{\mathrm B}}$. In order to highlight this striking observation, we take advantage of the long lifetime $\tau$ of dipolar excitons and define a large parameter of the theory
\begin{equation}
\label{SmallParameter}
\tilde \tau=\frac{\hbar q_0}{m\upsilon}\gg 1.
\end{equation}                 
In the thermodynamic limit $N_{\mathrm B}\rightarrow \infty$ we keep finite the combination $\sqrt{N_{\mathrm B}}/\tilde\tau$ by letting $\tau\rightarrow \infty $. This procedure eliminates the LT splitting from the few-body physics. Thus, the biexciton energy $\varepsilon$ and the wavefunction of the exciton relative motion $\phi_{\bm k}$ can be obtained from the one-channel Schrodinger equation
\begin{equation}
\left(-\frac{\hbar^2}{m}\Delta+V_{\uparrow\downarrow}(\bm r)\right)\varphi_{\uparrow\downarrow}(\bm r)=\varepsilon \varphi_{\uparrow\downarrow}(\bm r),
\end{equation}
with
\begin{equation}
\label{phi}
\varphi_{\uparrow\downarrow}(\bm r)=\frac{1}{\sqrt{S}}\sum_{\bm k}\phi_{\bm k} e^{i\bm k\bm r}.
\end{equation}
By virtue of the spherical symmetry of $V_{\uparrow\downarrow}(\bm r)$ the phase of $\phi_{\bm k}$ does not depend on $\bm k$. Our choice of the sign of the last term in \eqref{OmegaHamiltonian} corresponds to this phase being zero. Since we are concerned with the wavevectors $k$ much lower than the inverse molecular radius, we conveniently take $\phi_{\bm k}\equiv 1$. 

It is also convenient to define a small parameter 
\begin{equation}
\tilde\alpha=\frac{m\alpha}{\hbar^2 q_0^2}\ll1,
\end{equation}
which goes to zero as $S\rightarrow\infty$ in the thermodynamic limit. The combination $\sqrt{N_{\mathrm B}}\tilde\alpha$ is fixed by the biexciton density $N_{\mathrm B}/S$. We shall also keep fixed the product 
\begin{equation}
\label{Broadenings}
\tilde\alpha\tilde\tau\sim 1,
\end{equation}
which reads $\alpha\tau\sim\hbar$ in the original units.

In the $\bm \Omega$-XX phase one may neglect the interaction between unpaired excitons [the third term in the Hamiltonian \eqref{OmegaHamiltonian}]. The standard Bogoliubov-de-Gennes approach then yields the spectrum of elementary excitations
\begin{equation}
\label{ExcitationSpectrum}
\varepsilon_{\bm p}^{(\pm)}=\sqrt{\xi_{\bm p}^2-N_{\mathrm B}(\alpha\mp 2\hbar\upsilon p)^2}.
\end{equation}
In the interval
\begin{equation}
\label{RotonInterval}
2N_{\mathrm B}\tilde\tau^{-2}<\tilde\xi_0<4N_{\mathrm B}\tilde\tau^{-2}
\end{equation}
the lower branch of the spectrum develops a circle of roton-like minima at
\begin{equation}
\label{RotonPosition}
p_{\bm\Omega}=q_0\sqrt{2(4N_{\mathrm B}\tilde\tau^{-2}-\tilde\xi_0)}
\end{equation}
(see Fig. \ref{Fig1}). Here $\tilde\xi_0=m\xi_0/\hbar^2 q_0^2$ is a dimensionless measure of proximity to the second-order phase transition boundary in the radiative regime. The new gap reads
\begin{equation}
\label{RotonGap}
\Delta_{\bm\Omega}=\sqrt{8N_{\mathrm B}\tilde\tau^{-2}(\tilde\xi_0-2N_{\mathrm B}\tilde\tau^{-2})}
\end{equation}
and its closure defines a second-order transition to a novel $\bm{\Omega}$-X-XX phase whose nature will be elucidated below. In writing Eqs. \eqref{RotonInterval},\eqref{RotonPosition} and \eqref{RotonGap} we have taken advantage of the relationship \eqref{Broadenings}. In the $\varepsilon$ versus $\mu$ phase diagram the $\bm{\Omega}$-XX/$\bm{\Omega}$-X-XX boundary is obtained by counter-clock-wise rotation of the ray \eqref{XX/X-XX} substituting $g_{\mathrm X\mathrm B}$ by 
\begin{equation}
g_{\mathrm X\mathrm B}^\ast=g_{\mathrm X\mathrm B}-(2\pi) ^{-1}\frac{2\hbar^2}{m}(\tilde\alpha\tilde\tau)^{-2}\tilde\beta,
\end{equation}
with $\tilde\beta=m\beta/\hbar^2 q_0^2$. We thus see, that the exchange fields in combination with the natural broadening of the resonance $\beta$ favour the inequality $\sqrt{g_{\mathrm X} g_{\mathrm B}}>g_{\mathrm X\mathrm B}^\ast$ suggesting existence of a mixture of excitons and their molecules.

\begin{figure}
\includegraphics[width=1\linewidth]{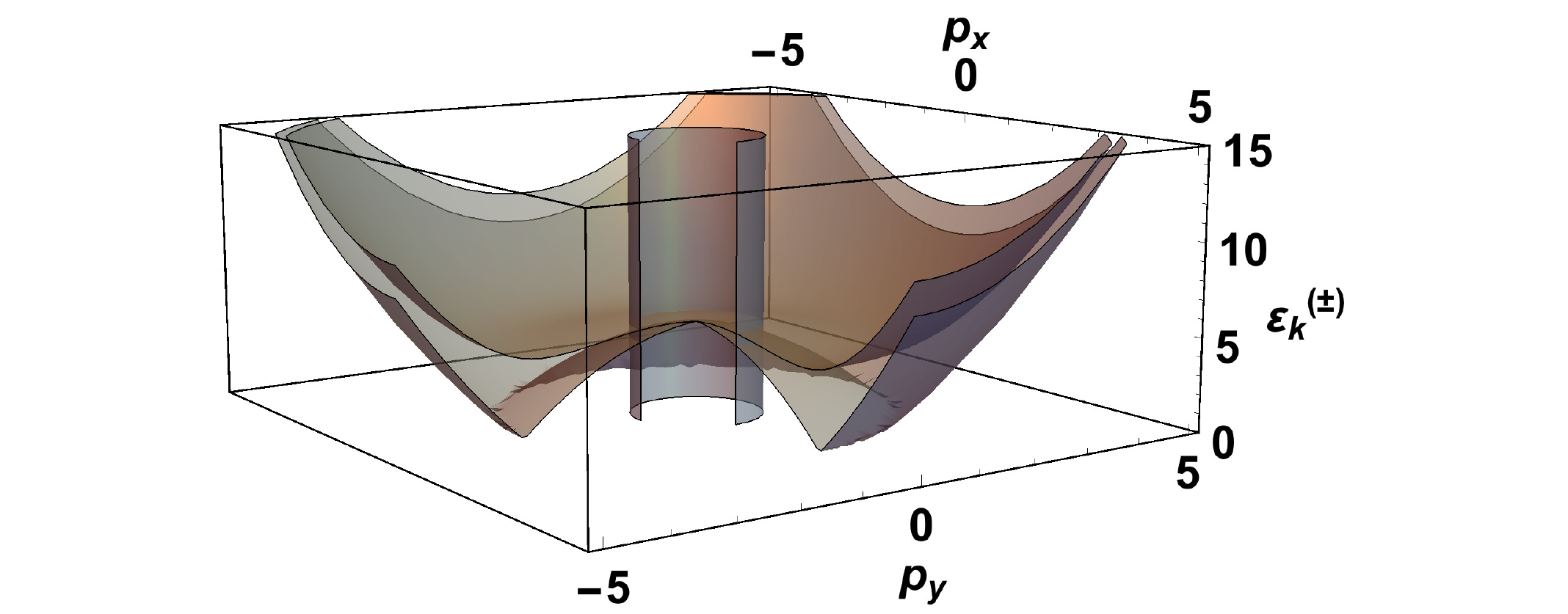} 
\caption{\label{Fig2}The pair-breaking excitation spectrum of the $\bm\Omega$-XX phase [Eq. \eqref{ExcitationSpectrum}]. We use the units $\hbar^2 q_0^2/m$ and $q_0$ for the energy and wavevectors, respectively. In these units we take $\tilde\xi_0=5$ and $\sqrt{N_{\mathrm B}}\tilde\alpha=0.5$. The collective effect of the exchange fields is governed by the combination $\sqrt{N_{\mathrm B}}/\tilde\tau$ and consists in appearance of a circle of roton-like minima in each branch of the spectrum. We take $\sqrt{N_{\mathrm B}}/\tilde\tau=1.5$, which corresponds to several hundreds of excitons in a typical TMD-based heterostructure \cite{Rivera2015}. The tube in the middle marks the boundary of the light cone.}
\end{figure}        

In contrast to the X-XX mixture discussed previously for the radiative regime, the X-component of the $\bm{\Omega}$-X-XX phase represents a superposition of plane-wave condensates. The corresponding mean-field energy density can be obtained from Eq. \eqref{OmegaHamiltonian} by substituting $\hat a_{\sigma, \bm p}/\sqrt{S}\rightarrow \Psi_{\sigma,\pm \bm q}\delta_{\bm p,\pm\bm q}$, where $\delta_{\bm p,\bm q}=1$ for $\bm p=\bm q$ and $\delta_{\bm p,\bm q}=0$ otherwise. Canonical transformation of the resulting quadratic form shows that the minimum of the energy is achieved by the choice $\Psi_{\sigma,\bm q}=\Psi_{-\sigma,-\bm q}^\ast$ and $\Psi_{\uparrow, \bm q}=-e^{-2i\varphi}\Psi_{\downarrow,\bm q}$ with $\bm q=\bm p_{\bm\Omega}$,   
\begin{equation}
\label{PlaneWaveVector}
p_{\bm\Omega}=2\sqrt{N_{\mathrm B}}\tilde\tau^{-1}q_0
\end{equation}
and $\lvert \Psi_{\uparrow,\pm\bm q}\rvert=\lvert \Psi_{\downarrow,\pm\bm q}\rvert\equiv \lvert \Psi_{\mathrm X}\rvert$, where
\begin{equation}
\lvert \Psi_{\mathrm X}\rvert=\sqrt{\frac{2N_{\mathrm B}m\upsilon^2+\sqrt{N_{\mathrm B}}\alpha+\mu-g_{\mathrm X\mathrm B}\lvert \Psi_{\mathrm B}\rvert^2}{2g_{\mathrm X}}}.
\end{equation}
The exciton order parameter
\begin{equation}
\label{PsiX}
\Psi_{\mathrm X}(\bm r)=\tfrac{1}{2}(-e^{-2i\varphi}\ket{\uparrow}+\ket{\downarrow})(\Psi_{\mathrm X}e^{i\bm p_{\bm\Omega}\bm r}-e^{2i\varphi}\Psi_{\mathrm X}^\ast e^{-i\bm p_{\bm\Omega}\bm r})
\end{equation}
is a mixture of counter-propagating plane-wave exciton condensates with their spins aligned along the exchange fields $\bm{\Omega}(\pm\bm p_{\bm\Omega})$ (all pointing in the same direction). Such spin configuration corresponds to the exciton linear polarization perpendicular to $\bm p_{\bm\Omega}$, consistently with the transverse polarization of the lower eigenstate of \eqref{H0}. As one reduces the molecular fraction $N_{\mathrm B}$, the wavevector \eqref{PlaneWaveVector} approaches the light cone. At that point the $\bm{\Omega}$-X-XX phase merges with the radiative exciton condensate X. Detailed investigation of this latter phase transition is outside the scope of the present study and will be given elsewhere.

\textit{Conclusions and outlook.} Our consideration reveals that the molecular coherence at the BCS-BEC phase transition induces spontaneous alignment of (pseudo)spins and effective magnetic fields associated with the constituent particles. In contrast to the phenomenon of SO-coupled BEC \cite{Galitski, Stringari, Ketterle, Hanle, Spielman}, this purely many-body effect does not require the presence of a finite-momentum minimum in the single-particle dispersion. The effect is suppressed on the BEC side of the transition by the mean-field repulsive interactions. A roton-like feature in the pair-breaking excitation spectrum of the molecular BEC may be considered as a precursor of the effect. The proposed setting of bright dipolar excitons in TMD's and QW's provides the most neat demonstration of the new phenomena due to smallness of the LT splitting of the single-exciton states and monotonous character of their dispersions. However, our finding clearly goes far beyond this particular scenario. We expect that our idea may be extended to the problems of fermionic pairing in the presence of SO-coupling \cite{SOCFermiSuperfluids}. Thus, our approach may shed light on the mechanism of formation of polarized excitonic condensates from SO-coupled electrons and holes \cite{Hanle}, on the influence of SO coupling on the BCS-BEC crossover \cite{Leggett, Hanai2017} and on the occurrence of line nodes in the excitation gap of (otherwise) conventional s-wave superconductors \cite{Samokhin}.

\textit{Acknowledgements.} The author acknowledges the support by Russian Science Foundation (Grant No. 18-72-00013).

\section*{Methods}

In order to derive the SO-coupled term appearing in the BCS-like Hamiltonian \eqref{OmegaHamiltonian}, we first consider single-particle Hamiltonians of the form
\begin{equation}
\label{SinglePolariton}
\hat H_i=\frac{\hbar^2}{2}\left(\begin{array}{cc}
m_\ast^{-1}\bm {\hat k}_i^2&m_\mathrm{LT}^{-1}\bm {\hat k}_{-,i}^2\\
m_\mathrm{LT}^{-1}\bm {\hat k}_{+,i}^2&m_\ast^{-1}\bm {\hat k}_i^2
\end{array}\right)
\end{equation} 
which posses the useful property
\begin{equation}
\label{TwoBody}
\begin{split}
&\hat H_1+\hat H_2=\\
&\frac{\hbar^2}{4}\left(\begin{array}{cc}
m_\ast^{-1}\bm {\hat K}^2&m_\mathrm{LT}^{-1}\bm {\hat K}_{-}^2\\
m_\mathrm{LT}^{-1}\bm {\hat K}_{+}^2&m_\ast^{-1}\bm {\hat K}^2
\end{array}\right)+
\hbar^2\left(\begin{array}{cc}
m_\ast^{-1}\bm {\hat k}^2&m_\mathrm{LT}^{-1}\bm {\hat k}_{-}^2\\
m_\mathrm{LT}^{-1}\bm {\hat k}_{+}^2&m_\ast^{-1}\bm {\hat k}^2
\end{array}\right).
\end{split}
\end{equation}
Here the index $i$ labels the particles,
\begin{equation*}
 m_\ast=m+m_\mathrm{LT}
 \end{equation*}
with $m$ being the particle mass, $\bm {\hat k}_{\pm, i}=\hat k_{x,i}\pm i\hat k_{y,i}$ and we have introduced $\bm{\hat K}=\bm {\hat k}_1+\bm {\hat k}_2$, $\bm {\hat K}_{\pm}=\bm {\hat k}_{\pm,1}+\bm {\hat k}_{\pm,2}$, $\bm{\hat k}=(\bm {\hat k}_1-\bm {\hat k}_2)/2$, $\bm {\hat k}_{\pm}=(\bm {\hat k}_{\pm,1}-\bm {\hat k}_{\pm,2})/2$. The spin basis is $\ket{\uparrow}$ and $\ket{\downarrow}$. The Hamiltonians of the type \eqref{SinglePolariton} govern the dynamics of the lower exciton-polaritons in planar semiconductor microcavities. Derivation of a BCS-like Hamiltonian for the single-particle Hamiltonians \eqref{SinglePolariton} will be presented below. The BCS-like pairing Hamiltonian \eqref{OmegaHamiltonian} for the $\bm \Omega$-X-XX and $\bm \Omega$-XX phases of excitons, which is of interest in this work, can be obtained along the same lines in the particular case of zero center-of-mass momentum of the pairs. The corresponding passage will be discussed in the end of the section.

In the second quantization the general form of the pairing Hamiltonian reads
\begin{equation}
\label{Hpair}
\hat H=\frac{1}{2}\sum_{\sigma_1\sigma_2\sigma_3\sigma_4}\sum_{\bm k,\bm K,\bm q,\bm Q}\mathcal H_{\sigma_1\sigma_2\sigma_3\sigma_4}^{\bm k,\bm K,\bm q,\bm Q}\hat C_{\sigma_2\sigma_1, \bm q,\bm Q}^{\dag} \hat C_{\sigma_3\sigma_4, \bm k,\bm K },
\end{equation} 
where the pair annihilation operator $\hat C_{\sigma\sigma^\prime, \bm k,\bm K }$ stands either for a pair of free particles with the same spin, $\hat C_{\sigma\sigma, \bm k,\bm K }\equiv\hat a_{\sigma,-\bm k+\bm K/2}\hat a_{\sigma,\bm k+\bm K/2}$, or for the molecular operator, $\hat C_{\uparrow\downarrow, \bm k,\bm K }=\hat C_{\downarrow\uparrow, \bm k,\bm K }\equiv \hat B_{\bm K}$. In the latter case the summation in \eqref{Hpair} over $\bm k$ ($\bm q$) is consistently absent. The matrix elements are defined as
\begin{equation}
\label{MatrixElem}
\mathcal H_{\sigma_1\sigma_2\sigma_3\sigma_4}^{\bm k,\bm K,\bm q,\bm Q}=\int \psi_{\sigma_2\sigma_1}^\ast(\bm r,\bm R)\mathcal {\hat H} \psi_{\sigma_3\sigma_4}(\bm r,\bm R)d\bm r d\bm R
\end{equation} 
with $\bm r=\bm r_1-\bm r_2$ and $\bm R=(\bm r_1+\bm r_2)/2$. The wave functions are taken as
\begin{equation} 
\psi_{\sigma\sigma}(\bm r,\bm R)=\frac{1}{S}e^{i\bm k\bm r+i\bm K\bm R}\ket{\sigma\sigma}
\end{equation}
for free particles and
\begin{equation}
\psi_{\uparrow\downarrow}(\bm r,\bm R)=\frac{1}{\sqrt{S}}e^{i\bm K\bm R}\varphi_{\uparrow\downarrow}(\bm r)\ket{\uparrow\downarrow}
\end{equation}
for the molecules. Here, the function $\varphi_{\uparrow\downarrow}(\bm r)$ is a solution of the Schrodinger equation
\begin{equation}
\label{RelativeMotion}
\left (\frac{\hbar^2 \hat {\bm k}^2}{m_\ast}+V_{\uparrow\downarrow}(r)\right)\varphi_{\uparrow\downarrow}(\bm r)=\varepsilon \varphi_{\uparrow\downarrow}(\bm r),
\end{equation}
where the 2D potential $V_{\uparrow\downarrow}(r)$ is assumed to be spherically symmetric. Upon substitution of
\begin{equation}
\label{phi}
\varphi_{\uparrow\downarrow}(\bm r)=\frac{1}{\sqrt{S}}\sum_{\bm k}\phi_{\bm k} e^{i\bm k\bm r}
\end{equation}
into Eq. \eqref{RelativeMotion} one gets
\begin{equation}
\left(\varepsilon-\frac{\hbar^2 k^2}{m_\ast}\right)\phi_{\bm k}=\frac{1}{S}\sum_{\bm q}V_{\uparrow\downarrow}(\bm q-\bm k)\phi_{\bm q},
\end{equation}
which shows that \textit{the phase} of the function $\phi_{\bm k}$ does not depend on $\bm k$. 

The operator $\mathcal {\hat H}$ in Eq. \eqref{MatrixElem} is a sum of a free part constructed by the Kronecker summation of the single-particle Hamiltonians \eqref{SinglePolariton} and the two-body interaction:
\begin{equation}
\mathcal {\hat H}=\hat H_1\oplus \hat H_2+\mathcal{\hat V},
\end{equation}
where
\begin{equation}
\mathcal{\hat V}=\left(\begin{array}{cccc}
V_{\uparrow\uparrow}(\bm r)&0&0&0\\
0&V_{\uparrow\downarrow}(\bm r)&0&0\\
0&0&V_{\downarrow\uparrow}(\bm r)&0\\
0&0&0&V_{\downarrow\downarrow}(\bm r)
\end{array}\right).
\end{equation}

The diagonal contribution has three terms. The first one
\begin{equation*}
\frac{1}{2}\sum_{\sigma,\bm k, \bm K}\left(\frac{\hbar^2\bm K^2}{4m_\ast}+\frac{\hbar^2 \bm k^2}{m_\ast}\right)\hat C_{\sigma\sigma, \bm k,\bm K }^\dagger\hat C_{\sigma\sigma, \bm k,\bm K }
\end{equation*}         
is just the overall kinetic energy in a subsystem of free particles counted by pairs. This term can be recast in the standard form
\begin{equation*}
\sum_{\sigma,\bm k}\frac{\hbar^2 \bm k^2}{2m_\ast}\hat a_{\sigma,\bm k}^\dagger\hat a_{\sigma,\bm k}.
\end{equation*}
The second term is the two-body interaction between the particles with alike spins:
\begin{equation*}
\frac{1}{2S}\sum_{\bm k,\bm q, \bm K,\sigma} \hat a_{\sigma, -\bm q+\bm K/2}^\dagger \hat a_{\sigma,\bm q+\bm K/2}^\dagger V_{\sigma \sigma}(\bm k-\bm q)\hat a_{\sigma, -\bm k+\bm K/2}\hat a_{\sigma,\bm k+\bm K/2}.
\end{equation*}
This term can be approximated by \cite{Utesov}
\begin{equation*} 
\frac{g}{2S}\sum_{\bm k,\bm q, \bm K,\sigma} \hat a_{\sigma, -\bm q+\bm K/2}^\dagger \hat a_{\sigma,\bm q+\bm K/2}^\dagger \hat a_{\sigma, -\bm k+\bm K/2}\hat a_{\sigma,\bm k+\bm K/2},
\end{equation*}
where $g$ is the effective potential and we omit the momentum-dependent correction due to the dipolar tail of the bair potential assuming that this correction is small as compared to the contact part. This excludes the dipolar-supersolid scenario from the consideration \cite{Andreev2015, RP, Fil2016}.

The last diagonal term
\begin{equation*}
\sum_{\bm K}\left(\frac{\hbar^2\bm K^2}{4m_\ast}+\varepsilon\right)\hat B_{\bm K}^\dagger \hat B_{\bm K} 
\end{equation*}
is the energy of free molecules. 

By collecting the off-diagonal terms one obtains
\begin{widetext}
\begin{equation*}
\frac{\hbar^2}{m_\mathrm{LT}}\sum_{\bm k, \bm K}\left[\left(\frac{\bm K_{-}^2}{4}+\bm k_{-}^2\right)\phi_{\bm k}\hat B_{\bm K}\hat a_{\uparrow, -\bm k+\bm K/2}^\dagger \hat a_{\uparrow,\bm k+\bm K/2}^\dagger+\left(\frac{\bm K_{+}^2}{4}+\bm k_{+}^2\right)\phi_{\bm k}\hat B_{\bm K}\hat a_{\downarrow, -\bm k+\bm K/2}^\dagger \hat a_{\downarrow,\bm k+\bm K/2}^\dagger+\mathrm{H.c.}\right].    
\end{equation*}
\end{widetext}
The molecular condensate is characterized by macroscopic occupation of the molecular state with $\bm K=0$. Following the Bogoliubov prescription one may let $\hat B_{\bm K}=\sqrt{N_{\mathrm B}}\delta_{\bm K0}$, where $\delta_{\bm K0}=1$ for $\bm K=0$ and $\delta_{\bm K0}\equiv0$ otherwise. One may then see, that the pair coherence enhances dramatically the effective magnetic field due to the LT splitting. For large molecular occupation number $N_{\mathrm B}\gg 1$ one may have a sizeable effect even for vanishingly small LT splitting of the single-particle states. The use of a single Eq. \eqref{RelativeMotion} (where one may let $m_\ast=m$) for determination of the bound state energy and wavefunction is fully justified in this case.

Finally, we notice that the requirement for the single-particle Hamiltonians of the type \eqref{SinglePolariton} to posses the property \eqref{TwoBody} may be relaxed if in considering the molecules one limits themselves to the $\bm K=0$ state only. Thus, for a radiative doublet of excitons with $k>q_0$, where $q_0$ is the boundary of the "light cone", one may write
\begin{widetext}
\begin{equation}
\begin{split}
\hat H&=\sum_{\sigma,\bm k}\left(\frac{\hbar^2 \bm k^2}{2m}+\frac{\hbar\upsilon k}{2}\right)\hat a_{\sigma,\bm k}^\dagger\hat a_{\sigma,\bm k}+\frac{g}{2S}\sum_{\bm k,\bm q, \bm K,\sigma} \hat a_{\sigma, -\bm q+\bm K/2}^\dagger \hat a_{\sigma,\bm q+\bm K/2}^\dagger \hat a_{\sigma, -\bm k+\bm K/2}\hat a_{\sigma,\bm k+\bm K/2}\\
&+\varepsilon N_{\mathrm B}+\hbar\upsilon\sqrt{N_{\mathrm B}}\sum_{\bm k}\left(k e^{-2i\varphi}\phi_{\bm k}\hat a_{\uparrow, -\bm k}^\dagger \hat a_{\uparrow,\bm k}^\dagger+ke^{2i\varphi}\phi_{\bm k} \hat a_{\downarrow, -\bm k}^\dagger \hat a_{\downarrow,\bm k}^\dagger+\mathrm{H.c.}\right).
\end{split}
\end{equation}
\end{widetext}
In the thermodynamic limit $N_{\mathrm B}\rightarrow \infty$ one should let $\upsilon\rightarrow 0$, omit the $\hbar\upsilon k/2$  term in the kinetic energy and use Eq. \eqref{RelativeMotion} with $m_\ast=m$ for determination of $\varepsilon$ and $\phi_{\bm k}$.

\bibliography{References}

\begin{thebibliography}{43}%
\makeatletter
\providecommand \@ifxundefined [1]{%
 \@ifx{#1\undefined}
}%
\providecommand \@ifnum [1]{%
 \ifnum #1\expandafter \@firstoftwo
 \else \expandafter \@secondoftwo
 \fi
}%
\providecommand \@ifx [1]{%
 \ifx #1\expandafter \@firstoftwo
 \else \expandafter \@secondoftwo
 \fi
}%
\providecommand \natexlab [1]{#1}%
\providecommand \enquote  [1]{``#1''}%
\providecommand \bibnamefont  [1]{#1}%
\providecommand \bibfnamefont [1]{#1}%
\providecommand \citenamefont [1]{#1}%
\providecommand \href@noop [0]{\@secondoftwo}%
\providecommand \href [0]{\begingroup \@sanitize@url \@href}%
\providecommand \@href[1]{\@@startlink{#1}\@@href}%
\providecommand \@@href[1]{\endgroup#1\@@endlink}%
\providecommand \@sanitize@url [0]{\catcode `\\12\catcode `\$12\catcode
  `\&12\catcode `\#12\catcode `\^12\catcode `\_12\catcode `\%12\relax}%
\providecommand \@@startlink[1]{}%
\providecommand \@@endlink[0]{}%
\providecommand \url  [0]{\begingroup\@sanitize@url \@url }%
\providecommand \@url [1]{\endgroup\@href {#1}{\urlprefix }}%
\providecommand \urlprefix  [0]{URL }%
\providecommand \Eprint [0]{\href }%
\providecommand \doibase [0]{http://dx.doi.org/}%
\providecommand \selectlanguage [0]{\@gobble}%
\providecommand \bibinfo  [0]{\@secondoftwo}%
\providecommand \bibfield  [0]{\@secondoftwo}%
\providecommand \translation [1]{[#1]}%
\providecommand \BibitemOpen [0]{}%
\providecommand \bibitemStop [0]{}%
\providecommand \bibitemNoStop [0]{.\EOS\space}%
\providecommand \EOS [0]{\spacefactor3000\relax}%
\providecommand \BibitemShut  [1]{\csname bibitem#1\endcsname}%
\let\auto@bib@innerbib\@empty
%</preamble>
\bibitem [{\citenamefont {Hanamura}\ and\ \citenamefont
  {Haug}(1977)}]{Hanamura}%
  \BibitemOpen
  \bibfield  {author} {\bibinfo {author} {\bibfnamefont {E.}~\bibnamefont
  {Hanamura}}\ and\ \bibinfo {author} {\bibfnamefont {H.}~\bibnamefont
  {Haug}},\ }\href {\doibase https://doi.org/10.1016/0370-1573(77)90012-6}
  {\bibfield  {journal} {\bibinfo  {journal} {Physics Reports}\ }\textbf
  {\bibinfo {volume} {33}},\ \bibinfo {pages} {209 } (\bibinfo {year}
  {1977})}\BibitemShut {NoStop}%
\bibitem [{\citenamefont {Moskalenko}(1958)}]{Moskalenko}%
  \BibitemOpen
  \bibfield  {author} {\bibinfo {author} {\bibfnamefont {S.}~\bibnamefont
  {Moskalenko}},\ }\href@noop {} {\bibfield  {journal} {\bibinfo  {journal}
  {Zh. Opt. Spektrosk.}\ }\textbf {\bibinfo {volume} {5}},\ \bibinfo {pages}
  {147} (\bibinfo {year} {1958})}\BibitemShut {NoStop}%
\bibitem [{\citenamefont {Lampert}(1958)}]{Lampert}%
  \BibitemOpen
  \bibfield  {author} {\bibinfo {author} {\bibfnamefont {M.~A.}\ \bibnamefont
  {Lampert}},\ }\href {\doibase 10.1103/PhysRevLett.1.450} {\bibfield
  {journal} {\bibinfo  {journal} {Phys. Rev. Lett.}\ }\textbf {\bibinfo
  {volume} {1}},\ \bibinfo {pages} {450} (\bibinfo {year} {1958})}\BibitemShut
  {NoStop}%
\bibitem [{\citenamefont {Mysyrowicz}\ \emph {et~al.}(1968)\citenamefont
  {Mysyrowicz}, \citenamefont {Grun}, \citenamefont {Levy}, \citenamefont
  {Bivas},\ and\ \citenamefont {Nikitine}}]{Nikitine}%
  \BibitemOpen
  \bibfield  {author} {\bibinfo {author} {\bibfnamefont {A.}~\bibnamefont
  {Mysyrowicz}}, \bibinfo {author} {\bibfnamefont {J.}~\bibnamefont {Grun}},
  \bibinfo {author} {\bibfnamefont {R.}~\bibnamefont {Levy}}, \bibinfo {author}
  {\bibfnamefont {A.}~\bibnamefont {Bivas}}, \ and\ \bibinfo {author}
  {\bibfnamefont {S.}~\bibnamefont {Nikitine}},\ }\href {\doibase
  https://doi.org/10.1016/0375-9601(68)90154-0} {\bibfield  {journal} {\bibinfo
   {journal} {Physics Letters A}\ }\textbf {\bibinfo {volume} {26}},\ \bibinfo
  {pages} {615 } (\bibinfo {year} {1968})}\BibitemShut {NoStop}%
\bibitem [{\citenamefont {Agranovich}\ \emph {et~al.}(1971)\citenamefont
  {Agranovich}, \citenamefont {Efremov},\ and\ \citenamefont
  {Kaminskaya}}]{Agranovich}%
  \BibitemOpen
  \bibfield  {author} {\bibinfo {author} {\bibfnamefont {V.}~\bibnamefont
  {Agranovich}}, \bibinfo {author} {\bibfnamefont {N.}~\bibnamefont {Efremov}},
  \ and\ \bibinfo {author} {\bibfnamefont {H.}~\bibnamefont {Kaminskaya}},\
  }\href {\doibase https://doi.org/10.1016/0030-4018(71)90262-8} {\bibfield
  {journal} {\bibinfo  {journal} {Optics Communications}\ }\textbf {\bibinfo
  {volume} {3}},\ \bibinfo {pages} {387 } (\bibinfo {year} {1971})}\BibitemShut
  {NoStop}%
\bibitem [{\citenamefont {Brinkman}\ \emph {et~al.}(1973)\citenamefont
  {Brinkman}, \citenamefont {Rice},\ and\ \citenamefont {Bell}}]{Brinkman1}%
  \BibitemOpen
  \bibfield  {author} {\bibinfo {author} {\bibfnamefont {W.~F.}\ \bibnamefont
  {Brinkman}}, \bibinfo {author} {\bibfnamefont {T.~M.}\ \bibnamefont {Rice}},
  \ and\ \bibinfo {author} {\bibfnamefont {B.}~\bibnamefont {Bell}},\ }\href
  {\doibase 10.1103/PhysRevB.8.1570} {\bibfield  {journal} {\bibinfo  {journal}
  {Phys. Rev. B}\ }\textbf {\bibinfo {volume} {8}},\ \bibinfo {pages} {1570}
  (\bibinfo {year} {1973})}\BibitemShut {NoStop}%
\bibitem [{\citenamefont {Kulakovskii}\ \emph {et~al.}(1985)\citenamefont
  {Kulakovskii}, \citenamefont {Lysenko},\ and\ \citenamefont
  {Timofeev}}]{Kulakovskii}%
  \BibitemOpen
  \bibfield  {author} {\bibinfo {author} {\bibfnamefont {V.~D.}\ \bibnamefont
  {Kulakovskii}}, \bibinfo {author} {\bibfnamefont {V.~G.}\ \bibnamefont
  {Lysenko}}, \ and\ \bibinfo {author} {\bibfnamefont {V.~B.}\ \bibnamefont
  {Timofeev}},\ }\href {\doibase 10.1070/PU1985v028n09ABEH003990} {\bibfield
  {journal} {\bibinfo  {journal} {Phys. Usp.}\ }\textbf {\bibinfo {volume}
  {28}},\ \bibinfo {pages} {735} (\bibinfo {year} {1985})}\BibitemShut
  {NoStop}%
\bibitem [{\citenamefont {{Nozi\`eres, P.}}\ and\ \citenamefont {{Saint James,
  D.}}(1982)}]{Noziers}%
  \BibitemOpen
  \bibfield  {author} {\bibinfo {author} {\bibnamefont {{Nozi\`eres, P.}}}\
  and\ \bibinfo {author} {\bibnamefont {{Saint James, D.}}},\ }\href {\doibase
  10.1051/jphys:019820043070113300} {\bibfield  {journal} {\bibinfo  {journal}
  {J. Phys. France}\ }\textbf {\bibinfo {volume} {43}},\ \bibinfo {pages}
  {1133} (\bibinfo {year} {1982})}\BibitemShut {NoStop}%
\bibitem [{\citenamefont {Miller}\ \emph {et~al.}(1982)\citenamefont {Miller},
  \citenamefont {Kleinman}, \citenamefont {Gossard},\ and\ \citenamefont
  {Munteanu}}]{Miller}%
  \BibitemOpen
  \bibfield  {author} {\bibinfo {author} {\bibfnamefont {R.~C.}\ \bibnamefont
  {Miller}}, \bibinfo {author} {\bibfnamefont {D.~A.}\ \bibnamefont
  {Kleinman}}, \bibinfo {author} {\bibfnamefont {A.~C.}\ \bibnamefont
  {Gossard}}, \ and\ \bibinfo {author} {\bibfnamefont {O.}~\bibnamefont
  {Munteanu}},\ }\href {\doibase 10.1103/PhysRevB.25.6545} {\bibfield
  {journal} {\bibinfo  {journal} {Phys. Rev. B}\ }\textbf {\bibinfo {volume}
  {25}},\ \bibinfo {pages} {6545} (\bibinfo {year} {1982})}\BibitemShut
  {NoStop}%
\bibitem [{\citenamefont {Kim}\ and\ \citenamefont {Wolfe}(1998)}]{Wolfe}%
  \BibitemOpen
  \bibfield  {author} {\bibinfo {author} {\bibfnamefont {J.~C.}\ \bibnamefont
  {Kim}}\ and\ \bibinfo {author} {\bibfnamefont {J.~P.}\ \bibnamefont
  {Wolfe}},\ }\href {\doibase 10.1103/PhysRevB.57.9861} {\bibfield  {journal}
  {\bibinfo  {journal} {Phys. Rev. B}\ }\textbf {\bibinfo {volume} {57}},\
  \bibinfo {pages} {9861} (\bibinfo {year} {1998})}\BibitemShut {NoStop}%
\bibitem [{\citenamefont {Spiegel}\ \emph {et~al.}(1997)\citenamefont
  {Spiegel}, \citenamefont {Bacher}, \citenamefont {Forchel}, \citenamefont
  {Jobst}, \citenamefont {Hommel},\ and\ \citenamefont {Landwehr}}]{Spiegel}%
  \BibitemOpen
  \bibfield  {author} {\bibinfo {author} {\bibfnamefont {R.}~\bibnamefont
  {Spiegel}}, \bibinfo {author} {\bibfnamefont {G.}~\bibnamefont {Bacher}},
  \bibinfo {author} {\bibfnamefont {A.}~\bibnamefont {Forchel}}, \bibinfo
  {author} {\bibfnamefont {B.}~\bibnamefont {Jobst}}, \bibinfo {author}
  {\bibfnamefont {D.}~\bibnamefont {Hommel}}, \ and\ \bibinfo {author}
  {\bibfnamefont {G.}~\bibnamefont {Landwehr}},\ }\href {\doibase
  10.1103/PhysRevB.55.9866} {\bibfield  {journal} {\bibinfo  {journal} {Phys.
  Rev. B}\ }\textbf {\bibinfo {volume} {55}},\ \bibinfo {pages} {9866}
  (\bibinfo {year} {1997})}\BibitemShut {NoStop}%
\bibitem [{\citenamefont {You}\ \emph {et~al.}(2015)\citenamefont {You},
  \citenamefont {Zhang}, \citenamefont {Berkelbach}, \citenamefont {Hybertsen},
  \citenamefont {Reichman},\ and\ \citenamefont {Heinz}}]{XXinTMD1}%
  \BibitemOpen
  \bibfield  {author} {\bibinfo {author} {\bibfnamefont {Y.}~\bibnamefont
  {You}}, \bibinfo {author} {\bibfnamefont {X.-X.}\ \bibnamefont {Zhang}},
  \bibinfo {author} {\bibfnamefont {T.~C.}\ \bibnamefont {Berkelbach}},
  \bibinfo {author} {\bibfnamefont {M.~S.}\ \bibnamefont {Hybertsen}}, \bibinfo
  {author} {\bibfnamefont {D.~R.}\ \bibnamefont {Reichman}}, \ and\ \bibinfo
  {author} {\bibfnamefont {T.~F.}\ \bibnamefont {Heinz}},\ }\href {\doibase
  10.1038/nphys3324} {\bibfield  {journal} {\bibinfo  {journal} {Nature
  Physics}\ }\textbf {\bibinfo {volume} {11}},\ \bibinfo {pages} {477}
  (\bibinfo {year} {2015})}\BibitemShut {NoStop}%
\bibitem [{\citenamefont {Ye}\ \emph {et~al.}(2018)\citenamefont {Ye},
  \citenamefont {Waldecker}, \citenamefont {Ma}, \citenamefont {Rhodes},
  \citenamefont {Antony}, \citenamefont {Kim}, \citenamefont {Zhang},
  \citenamefont {Deng}, \citenamefont {Jiang}, \citenamefont {Lu},
  \citenamefont {Smirnov}, \citenamefont {Watanabe}, \citenamefont {Taniguchi},
  \citenamefont {Hone},\ and\ \citenamefont {Heinz}}]{XXinTMD2}%
  \BibitemOpen
  \bibfield  {author} {\bibinfo {author} {\bibfnamefont {Z.}~\bibnamefont
  {Ye}}, \bibinfo {author} {\bibfnamefont {L.}~\bibnamefont {Waldecker}},
  \bibinfo {author} {\bibfnamefont {E.~Y.}\ \bibnamefont {Ma}}, \bibinfo
  {author} {\bibfnamefont {D.}~\bibnamefont {Rhodes}}, \bibinfo {author}
  {\bibfnamefont {A.}~\bibnamefont {Antony}}, \bibinfo {author} {\bibfnamefont
  {B.}~\bibnamefont {Kim}}, \bibinfo {author} {\bibfnamefont {X.-X.}\
  \bibnamefont {Zhang}}, \bibinfo {author} {\bibfnamefont {M.}~\bibnamefont
  {Deng}}, \bibinfo {author} {\bibfnamefont {Y.}~\bibnamefont {Jiang}},
  \bibinfo {author} {\bibfnamefont {Z.}~\bibnamefont {Lu}}, \bibinfo {author}
  {\bibfnamefont {D.}~\bibnamefont {Smirnov}}, \bibinfo {author} {\bibfnamefont
  {K.}~\bibnamefont {Watanabe}}, \bibinfo {author} {\bibfnamefont
  {T.}~\bibnamefont {Taniguchi}}, \bibinfo {author} {\bibfnamefont
  {J.}~\bibnamefont {Hone}}, \ and\ \bibinfo {author} {\bibfnamefont {T.~F.}\
  \bibnamefont {Heinz}},\ }\href {\doibase 10.1038/s41467-018-05917-8}
  {\bibfield  {journal} {\bibinfo  {journal} {Nature Communications}\ }\textbf
  {\bibinfo {volume} {9}},\ \bibinfo {pages} {3718} (\bibinfo {year}
  {2018})}\BibitemShut {NoStop}%
\bibitem [{\citenamefont {Li}\ \emph {et~al.}(2018)\citenamefont {Li},
  \citenamefont {Wang}, \citenamefont {Lu}, \citenamefont {Jin}, \citenamefont
  {Chen}, \citenamefont {Meng}, \citenamefont {Lian}, \citenamefont
  {Taniguchi}, \citenamefont {Watanabe}, \citenamefont {Zhang}, \citenamefont
  {Smirnov},\ and\ \citenamefont {Shi}}]{XXinTMD3}%
  \BibitemOpen
  \bibfield  {author} {\bibinfo {author} {\bibfnamefont {Z.}~\bibnamefont
  {Li}}, \bibinfo {author} {\bibfnamefont {T.}~\bibnamefont {Wang}}, \bibinfo
  {author} {\bibfnamefont {Z.}~\bibnamefont {Lu}}, \bibinfo {author}
  {\bibfnamefont {C.}~\bibnamefont {Jin}}, \bibinfo {author} {\bibfnamefont
  {Y.}~\bibnamefont {Chen}}, \bibinfo {author} {\bibfnamefont {Y.}~\bibnamefont
  {Meng}}, \bibinfo {author} {\bibfnamefont {Z.}~\bibnamefont {Lian}}, \bibinfo
  {author} {\bibfnamefont {T.}~\bibnamefont {Taniguchi}}, \bibinfo {author}
  {\bibfnamefont {K.}~\bibnamefont {Watanabe}}, \bibinfo {author}
  {\bibfnamefont {S.}~\bibnamefont {Zhang}}, \bibinfo {author} {\bibfnamefont
  {D.}~\bibnamefont {Smirnov}}, \ and\ \bibinfo {author} {\bibfnamefont
  {S.-F.}\ \bibnamefont {Shi}},\ }\href {\doibase 10.1038/s41467-018-05863-5}
  {\bibfield  {journal} {\bibinfo  {journal} {Nature Communications}\ }\textbf
  {\bibinfo {volume} {9}},\ \bibinfo {pages} {3719} (\bibinfo {year}
  {2018})}\BibitemShut {NoStop}%
\bibitem [{\citenamefont {High}\ \emph {et~al.}(2012)\citenamefont {High},
  \citenamefont {Leonard}, \citenamefont {Hammack}, \citenamefont {Fogler},
  \citenamefont {Butov}, \citenamefont {Kavokin}, \citenamefont {Campman},\
  and\ \citenamefont {Gossard}}]{High2012}%
  \BibitemOpen
  \bibfield  {author} {\bibinfo {author} {\bibfnamefont {A.~A.}\ \bibnamefont
  {High}}, \bibinfo {author} {\bibfnamefont {J.~R.}\ \bibnamefont {Leonard}},
  \bibinfo {author} {\bibfnamefont {A.~T.}\ \bibnamefont {Hammack}}, \bibinfo
  {author} {\bibfnamefont {M.~M.}\ \bibnamefont {Fogler}}, \bibinfo {author}
  {\bibfnamefont {L.~V.}\ \bibnamefont {Butov}}, \bibinfo {author}
  {\bibfnamefont {A.~V.}\ \bibnamefont {Kavokin}}, \bibinfo {author}
  {\bibfnamefont {K.~L.}\ \bibnamefont {Campman}}, \ and\ \bibinfo {author}
  {\bibfnamefont {A.~C.}\ \bibnamefont {Gossard}},\ }\href {\doibase
  10.1038/nature10903} {\bibfield  {journal} {\bibinfo  {journal} {Nature}\
  }\textbf {\bibinfo {volume} {483}},\ \bibinfo {pages} {584} (\bibinfo {year}
  {2012})}\BibitemShut {NoStop}%
\bibitem [{\citenamefont {Anankine}\ \emph {et~al.}(2017)\citenamefont
  {Anankine}, \citenamefont {Beian}, \citenamefont {Dang}, \citenamefont
  {Alloing}, \citenamefont {Cambril}, \citenamefont {Merghem}, \citenamefont
  {Carbonell}, \citenamefont {Lema\^{\i}tre},\ and\ \citenamefont
  {Dubin}}]{Dubin}%
  \BibitemOpen
  \bibfield  {author} {\bibinfo {author} {\bibfnamefont {R.}~\bibnamefont
  {Anankine}}, \bibinfo {author} {\bibfnamefont {M.}~\bibnamefont {Beian}},
  \bibinfo {author} {\bibfnamefont {S.}~\bibnamefont {Dang}}, \bibinfo {author}
  {\bibfnamefont {M.}~\bibnamefont {Alloing}}, \bibinfo {author} {\bibfnamefont
  {E.}~\bibnamefont {Cambril}}, \bibinfo {author} {\bibfnamefont
  {K.}~\bibnamefont {Merghem}}, \bibinfo {author} {\bibfnamefont {C.~G.}\
  \bibnamefont {Carbonell}}, \bibinfo {author} {\bibfnamefont {A.}~\bibnamefont
  {Lema\^{\i}tre}}, \ and\ \bibinfo {author} {\bibfnamefont {F.~m.~c.}\
  \bibnamefont {Dubin}},\ }\href {\doibase 10.1103/PhysRevLett.118.127402}
  {\bibfield  {journal} {\bibinfo  {journal} {Phys. Rev. Lett.}\ }\textbf
  {\bibinfo {volume} {118}},\ \bibinfo {pages} {127402} (\bibinfo {year}
  {2017})}\BibitemShut {NoStop}%
\bibitem [{\citenamefont {Misra}\ \emph {et~al.}(2018)\citenamefont {Misra},
  \citenamefont {Stern}, \citenamefont {Joshua}, \citenamefont {Umansky},\ and\
  \citenamefont {Bar-Joseph}}]{BarJoseph}%
  \BibitemOpen
  \bibfield  {author} {\bibinfo {author} {\bibfnamefont {S.}~\bibnamefont
  {Misra}}, \bibinfo {author} {\bibfnamefont {M.}~\bibnamefont {Stern}},
  \bibinfo {author} {\bibfnamefont {A.}~\bibnamefont {Joshua}}, \bibinfo
  {author} {\bibfnamefont {V.}~\bibnamefont {Umansky}}, \ and\ \bibinfo
  {author} {\bibfnamefont {I.}~\bibnamefont {Bar-Joseph}},\ }\href {\doibase
  10.1103/PhysRevLett.120.047402} {\bibfield  {journal} {\bibinfo  {journal}
  {Phys. Rev. Lett.}\ }\textbf {\bibinfo {volume} {120}},\ \bibinfo {pages}
  {047402} (\bibinfo {year} {2018})}\BibitemShut {NoStop}%
\bibitem [{\citenamefont {Fogler}\ \emph {et~al.}(2014)\citenamefont {Fogler},
  \citenamefont {Butov},\ and\ \citenamefont {Novoselov}}]{Fogler2014}%
  \BibitemOpen
  \bibfield  {author} {\bibinfo {author} {\bibfnamefont {M.~M.}\ \bibnamefont
  {Fogler}}, \bibinfo {author} {\bibfnamefont {L.~V.}\ \bibnamefont {Butov}}, \
  and\ \bibinfo {author} {\bibfnamefont {K.~S.}\ \bibnamefont {Novoselov}},\
  }\href {\doibase 10.1038/ncomms5555} {\bibfield  {journal} {\bibinfo
  {journal} {Nature Communications}\ }\textbf {\bibinfo {volume} {5}},\
  \bibinfo {pages} {4555} (\bibinfo {year} {2014})}\BibitemShut {NoStop}%
\bibitem [{\citenamefont {Rivera}\ \emph {et~al.}(2015)\citenamefont {Rivera},
  \citenamefont {Schaibley}, \citenamefont {Jones}, \citenamefont {Ross},
  \citenamefont {Wu}, \citenamefont {Aivazian}, \citenamefont {Klement},
  \citenamefont {Seyler}, \citenamefont {Clark}, \citenamefont {Ghimire},
  \citenamefont {Yan}, \citenamefont {Mandrus}, \citenamefont {Yao},\ and\
  \citenamefont {Xu}}]{Rivera2015}%
  \BibitemOpen
  \bibfield  {author} {\bibinfo {author} {\bibfnamefont {P.}~\bibnamefont
  {Rivera}}, \bibinfo {author} {\bibfnamefont {J.~R.}\ \bibnamefont
  {Schaibley}}, \bibinfo {author} {\bibfnamefont {A.~M.}\ \bibnamefont
  {Jones}}, \bibinfo {author} {\bibfnamefont {J.~S.}\ \bibnamefont {Ross}},
  \bibinfo {author} {\bibfnamefont {S.}~\bibnamefont {Wu}}, \bibinfo {author}
  {\bibfnamefont {G.}~\bibnamefont {Aivazian}}, \bibinfo {author}
  {\bibfnamefont {P.}~\bibnamefont {Klement}}, \bibinfo {author} {\bibfnamefont
  {K.}~\bibnamefont {Seyler}}, \bibinfo {author} {\bibfnamefont
  {G.}~\bibnamefont {Clark}}, \bibinfo {author} {\bibfnamefont {N.~J.}\
  \bibnamefont {Ghimire}}, \bibinfo {author} {\bibfnamefont {J.}~\bibnamefont
  {Yan}}, \bibinfo {author} {\bibfnamefont {D.~G.}\ \bibnamefont {Mandrus}},
  \bibinfo {author} {\bibfnamefont {W.}~\bibnamefont {Yao}}, \ and\ \bibinfo
  {author} {\bibfnamefont {X.}~\bibnamefont {Xu}},\ }\href {\doibase
  10.1038/ncomms7242} {\bibfield  {journal} {\bibinfo  {journal} {Nature
  Communications}\ }\textbf {\bibinfo {volume} {6}},\ \bibinfo {pages} {6242}
  (\bibinfo {year} {2015})}\BibitemShut {NoStop}%
\bibitem [{\citenamefont {Wang}\ \emph {et~al.}(2018)\citenamefont {Wang},
  \citenamefont {Chernikov}, \citenamefont {Glazov}, \citenamefont {Heinz},
  \citenamefont {Marie}, \citenamefont {Amand},\ and\ \citenamefont
  {Urbaszek}}]{Colloquium}%
  \BibitemOpen
  \bibfield  {author} {\bibinfo {author} {\bibfnamefont {G.}~\bibnamefont
  {Wang}}, \bibinfo {author} {\bibfnamefont {A.}~\bibnamefont {Chernikov}},
  \bibinfo {author} {\bibfnamefont {M.~M.}\ \bibnamefont {Glazov}}, \bibinfo
  {author} {\bibfnamefont {T.~F.}\ \bibnamefont {Heinz}}, \bibinfo {author}
  {\bibfnamefont {X.}~\bibnamefont {Marie}}, \bibinfo {author} {\bibfnamefont
  {T.}~\bibnamefont {Amand}}, \ and\ \bibinfo {author} {\bibfnamefont
  {B.}~\bibnamefont {Urbaszek}},\ }\href {\doibase
  10.1103/RevModPhys.90.021001} {\bibfield  {journal} {\bibinfo  {journal}
  {Rev. Mod. Phys.}\ }\textbf {\bibinfo {volume} {90}},\ \bibinfo {pages}
  {021001} (\bibinfo {year} {2018})}\BibitemShut {NoStop}%
\bibitem [{\citenamefont {Schindler}\ and\ \citenamefont
  {Zimmermann}(2008)}]{Zimmerman}%
  \BibitemOpen
  \bibfield  {author} {\bibinfo {author} {\bibfnamefont {C.}~\bibnamefont
  {Schindler}}\ and\ \bibinfo {author} {\bibfnamefont {R.}~\bibnamefont
  {Zimmermann}},\ }\href {\doibase 10.1103/PhysRevB.78.045313} {\bibfield
  {journal} {\bibinfo  {journal} {Phys. Rev. B}\ }\textbf {\bibinfo {volume}
  {78}},\ \bibinfo {pages} {045313} (\bibinfo {year} {2008})}\BibitemShut
  {NoStop}%
\bibitem [{\citenamefont {Meyertholen}\ and\ \citenamefont
  {Fogler}(2008)}]{Fogler}%
  \BibitemOpen
  \bibfield  {author} {\bibinfo {author} {\bibfnamefont {A.~D.}\ \bibnamefont
  {Meyertholen}}\ and\ \bibinfo {author} {\bibfnamefont {M.~M.}\ \bibnamefont
  {Fogler}},\ }\href {\doibase 10.1103/PhysRevB.78.235307} {\bibfield
  {journal} {\bibinfo  {journal} {Phys. Rev. B}\ }\textbf {\bibinfo {volume}
  {78}},\ \bibinfo {pages} {235307} (\bibinfo {year} {2008})}\BibitemShut
  {NoStop}%
\bibitem [{\citenamefont {Andreev}(2015)}]{Andreev2015}%
  \BibitemOpen
  \bibfield  {author} {\bibinfo {author} {\bibfnamefont {S.~V.}\ \bibnamefont
  {Andreev}},\ }\href {\doibase 10.1103/PhysRevB.92.041117} {\bibfield
  {journal} {\bibinfo  {journal} {Phys. Rev. B}\ }\textbf {\bibinfo {volume}
  {92}},\ \bibinfo {pages} {041117(R)} (\bibinfo {year} {2015})}\BibitemShut
  {NoStop}%
\bibitem [{\citenamefont {Andreev}(2016)}]{RP}%
  \BibitemOpen
  \bibfield  {author} {\bibinfo {author} {\bibfnamefont {S.~V.}\ \bibnamefont
  {Andreev}},\ }\href {\doibase 10.1103/PhysRevB.94.140501} {\bibfield
  {journal} {\bibinfo  {journal} {Phys. Rev. B}\ }\textbf {\bibinfo {volume}
  {94}},\ \bibinfo {pages} {140501(R)} (\bibinfo {year} {2016})}\BibitemShut
  {NoStop}%
\bibitem [{\citenamefont {Gupalov}\ \emph {et~al.}(1998)\citenamefont
  {Gupalov}, \citenamefont {Ivchenko},\ and\ \citenamefont
  {Kavokin}}]{Gupalov}%
  \BibitemOpen
  \bibfield  {author} {\bibinfo {author} {\bibfnamefont {S.~V.}\ \bibnamefont
  {Gupalov}}, \bibinfo {author} {\bibfnamefont {E.~L.}\ \bibnamefont
  {Ivchenko}}, \ and\ \bibinfo {author} {\bibfnamefont {A.~V.}\ \bibnamefont
  {Kavokin}},\ }\href {\doibase 10.1134/1.558441} {\bibfield  {journal}
  {\bibinfo  {journal} {Journal of Experimental and Theoretical Physics}\
  }\textbf {\bibinfo {volume} {86}},\ \bibinfo {pages} {388} (\bibinfo {year}
  {1998})}\BibitemShut {NoStop}%
\bibitem [{\citenamefont {Glazov}\ \emph {et~al.}(2014)\citenamefont {Glazov},
  \citenamefont {Amand}, \citenamefont {Marie}, \citenamefont {Lagarde},
  \citenamefont {Bouet},\ and\ \citenamefont {Urbaszek}}]{Glazov}%
  \BibitemOpen
  \bibfield  {author} {\bibinfo {author} {\bibfnamefont {M.~M.}\ \bibnamefont
  {Glazov}}, \bibinfo {author} {\bibfnamefont {T.}~\bibnamefont {Amand}},
  \bibinfo {author} {\bibfnamefont {X.}~\bibnamefont {Marie}}, \bibinfo
  {author} {\bibfnamefont {D.}~\bibnamefont {Lagarde}}, \bibinfo {author}
  {\bibfnamefont {L.}~\bibnamefont {Bouet}}, \ and\ \bibinfo {author}
  {\bibfnamefont {B.}~\bibnamefont {Urbaszek}},\ }\href {\doibase
  10.1103/PhysRevB.89.201302} {\bibfield  {journal} {\bibinfo  {journal} {Phys.
  Rev. B}\ }\textbf {\bibinfo {volume} {89}},\ \bibinfo {pages} {201302(R)}
  (\bibinfo {year} {2014})}\BibitemShut {NoStop}%
\bibitem [{\citenamefont {Maialle}\ \emph {et~al.}(1993)\citenamefont
  {Maialle}, \citenamefont {de~Andrada~e Silva},\ and\ \citenamefont
  {Sham}}]{Sham}%
  \BibitemOpen
  \bibfield  {author} {\bibinfo {author} {\bibfnamefont {M.~Z.}\ \bibnamefont
  {Maialle}}, \bibinfo {author} {\bibfnamefont {E.~A.}\ \bibnamefont
  {de~Andrada~e Silva}}, \ and\ \bibinfo {author} {\bibfnamefont {L.~J.}\
  \bibnamefont {Sham}},\ }\href {\doibase 10.1103/PhysRevB.47.15776} {\bibfield
   {journal} {\bibinfo  {journal} {Phys. Rev. B}\ }\textbf {\bibinfo {volume}
  {47}},\ \bibinfo {pages} {15776} (\bibinfo {year} {1993})}\BibitemShut
  {NoStop}%
\bibitem [{\citenamefont {Dresselhaus}(1955)}]{Dresselhaus}%
  \BibitemOpen
  \bibfield  {author} {\bibinfo {author} {\bibfnamefont {G.}~\bibnamefont
  {Dresselhaus}},\ }\href {\doibase 10.1103/PhysRev.100.580} {\bibfield
  {journal} {\bibinfo  {journal} {Phys. Rev.}\ }\textbf {\bibinfo {volume}
  {100}},\ \bibinfo {pages} {580} (\bibinfo {year} {1955})}\BibitemShut
  {NoStop}%
\bibitem [{\citenamefont {Bychkov}\ and\ \citenamefont
  {Rashba}(1984)}]{BychkovRashba}%
  \BibitemOpen
  \bibfield  {author} {\bibinfo {author} {\bibfnamefont {Y.~A.}\ \bibnamefont
  {Bychkov}}\ and\ \bibinfo {author} {\bibfnamefont {E.~I.}\ \bibnamefont
  {Rashba}},\ }\href {\doibase 10.1088/0022-3719/17/33/015} {\bibfield
  {journal} {\bibinfo  {journal} {Journal of Physics C: Solid State Physics}\
  }\textbf {\bibinfo {volume} {17}},\ \bibinfo {pages} {6039} (\bibinfo {year}
  {1984})}\BibitemShut {NoStop}%
\bibitem [{\citenamefont {Dyakonov}\ and\ \citenamefont
  {Kachorovskii}(1986)}]{DyakonovKachorovskii}%
  \BibitemOpen
  \bibfield  {author} {\bibinfo {author} {\bibfnamefont {M.~I.}\ \bibnamefont
  {Dyakonov}}\ and\ \bibinfo {author} {\bibfnamefont {Y.~V.}\ \bibnamefont
  {Kachorovskii}},\ }\href@noop {} {\bibfield  {journal} {\bibinfo  {journal}
  {Sov. Phys. Semicond.}\ ,\ \bibinfo {pages} {110}} (\bibinfo {year}
  {1986})}\BibitemShut {NoStop}%
\bibitem [{\citenamefont {Stanescu}\ \emph {et~al.}(2008)\citenamefont
  {Stanescu}, \citenamefont {Anderson},\ and\ \citenamefont
  {Galitski}}]{Galitski}%
  \BibitemOpen
  \bibfield  {author} {\bibinfo {author} {\bibfnamefont {T.~D.}\ \bibnamefont
  {Stanescu}}, \bibinfo {author} {\bibfnamefont {B.}~\bibnamefont {Anderson}},
  \ and\ \bibinfo {author} {\bibfnamefont {V.}~\bibnamefont {Galitski}},\
  }\href {\doibase 10.1103/PhysRevA.78.023616} {\bibfield  {journal} {\bibinfo
  {journal} {Phys. Rev. A}\ }\textbf {\bibinfo {volume} {78}},\ \bibinfo
  {pages} {023616} (\bibinfo {year} {2008})}\BibitemShut {NoStop}%
\bibitem [{\citenamefont {Li}\ \emph {et~al.}(2012)\citenamefont {Li},
  \citenamefont {Pitaevskii},\ and\ \citenamefont {Stringari}}]{Stringari}%
  \BibitemOpen
  \bibfield  {author} {\bibinfo {author} {\bibfnamefont {Y.}~\bibnamefont
  {Li}}, \bibinfo {author} {\bibfnamefont {L.~P.}\ \bibnamefont {Pitaevskii}},
  \ and\ \bibinfo {author} {\bibfnamefont {S.}~\bibnamefont {Stringari}},\
  }\href {\doibase 10.1103/PhysRevLett.108.225301} {\bibfield  {journal}
  {\bibinfo  {journal} {Phys. Rev. Lett.}\ }\textbf {\bibinfo {volume} {108}},\
  \bibinfo {pages} {225301} (\bibinfo {year} {2012})}\BibitemShut {NoStop}%
\bibitem [{\citenamefont {Li}\ \emph {et~al.}(2017)\citenamefont {Li},
  \citenamefont {Lee}, \citenamefont {Huang}, \citenamefont {Burchesky},
  \citenamefont {Shteynas}, \citenamefont {Top}, \citenamefont {Jamison},\ and\
  \citenamefont {Ketterle}}]{Ketterle}%
  \BibitemOpen
  \bibfield  {author} {\bibinfo {author} {\bibfnamefont {J.-R.}\ \bibnamefont
  {Li}}, \bibinfo {author} {\bibfnamefont {J.}~\bibnamefont {Lee}}, \bibinfo
  {author} {\bibfnamefont {W.}~\bibnamefont {Huang}}, \bibinfo {author}
  {\bibfnamefont {S.}~\bibnamefont {Burchesky}}, \bibinfo {author}
  {\bibfnamefont {B.}~\bibnamefont {Shteynas}}, \bibinfo {author}
  {\bibfnamefont {F.~{\c C}.}\ \bibnamefont {Top}}, \bibinfo {author}
  {\bibfnamefont {A.~O.}\ \bibnamefont {Jamison}}, \ and\ \bibinfo {author}
  {\bibfnamefont {W.}~\bibnamefont {Ketterle}},\ }\href {\doibase
  10.1038/nature21431} {\bibfield  {journal} {\bibinfo  {journal} {Nature}\
  }\textbf {\bibinfo {volume} {543}},\ \bibinfo {pages} {91} (\bibinfo {year}
  {2017})}\BibitemShut {NoStop}%
\bibitem [{\citenamefont {Andreev}\ and\ \citenamefont
  {Nalitov}(2018)}]{Hanle}%
  \BibitemOpen
  \bibfield  {author} {\bibinfo {author} {\bibfnamefont {S.~V.}\ \bibnamefont
  {Andreev}}\ and\ \bibinfo {author} {\bibfnamefont {A.~V.}\ \bibnamefont
  {Nalitov}},\ }\href {\doibase 10.1103/PhysRevB.97.165139} {\bibfield
  {journal} {\bibinfo  {journal} {Phys. Rev. B}\ }\textbf {\bibinfo {volume}
  {97}},\ \bibinfo {pages} {165139} (\bibinfo {year} {2018})}\BibitemShut
  {NoStop}%
\bibitem [{\citenamefont {Putra}\ \emph {et~al.}(2020)\citenamefont {Putra},
  \citenamefont {Salces-C\'arcoba}, \citenamefont {Yue}, \citenamefont
  {Sugawa},\ and\ \citenamefont {Spielman}}]{Spielman}%
  \BibitemOpen
  \bibfield  {author} {\bibinfo {author} {\bibfnamefont {A.}~\bibnamefont
  {Putra}}, \bibinfo {author} {\bibfnamefont {F.}~\bibnamefont
  {Salces-C\'arcoba}}, \bibinfo {author} {\bibfnamefont {Y.}~\bibnamefont
  {Yue}}, \bibinfo {author} {\bibfnamefont {S.}~\bibnamefont {Sugawa}}, \ and\
  \bibinfo {author} {\bibfnamefont {I.~B.}\ \bibnamefont {Spielman}},\ }\href
  {\doibase 10.1103/PhysRevLett.124.053605} {\bibfield  {journal} {\bibinfo
  {journal} {Phys. Rev. Lett.}\ }\textbf {\bibinfo {volume} {124}},\ \bibinfo
  {pages} {053605} (\bibinfo {year} {2020})}\BibitemShut {NoStop}%
\bibitem [{\citenamefont {Utesov}\ \emph {et~al.}(2018)\citenamefont {Utesov},
  \citenamefont {Baglay},\ and\ \citenamefont {Andreev}}]{Utesov}%
  \BibitemOpen
  \bibfield  {author} {\bibinfo {author} {\bibfnamefont {O.~I.}\ \bibnamefont
  {Utesov}}, \bibinfo {author} {\bibfnamefont {M.~I.}\ \bibnamefont {Baglay}},
  \ and\ \bibinfo {author} {\bibfnamefont {S.~V.}\ \bibnamefont {Andreev}},\
  }\href {\doibase 10.1103/PhysRevA.97.053617} {\bibfield  {journal} {\bibinfo
  {journal} {Phys. Rev. A}\ }\textbf {\bibinfo {volume} {97}},\ \bibinfo
  {pages} {053617} (\bibinfo {year} {2018})}\BibitemShut {NoStop}%
\bibitem [{\citenamefont {Safonov}\ \emph {et~al.}(1998)\citenamefont
  {Safonov}, \citenamefont {Vasilyev}, \citenamefont {Yasnikov}, \citenamefont
  {Lukashevich},\ and\ \citenamefont {Jaakkola}}]{Safonov}%
  \BibitemOpen
  \bibfield  {author} {\bibinfo {author} {\bibfnamefont {A.~I.}\ \bibnamefont
  {Safonov}}, \bibinfo {author} {\bibfnamefont {S.~A.}\ \bibnamefont
  {Vasilyev}}, \bibinfo {author} {\bibfnamefont {I.~S.}\ \bibnamefont
  {Yasnikov}}, \bibinfo {author} {\bibfnamefont {I.~I.}\ \bibnamefont
  {Lukashevich}}, \ and\ \bibinfo {author} {\bibfnamefont {S.}~\bibnamefont
  {Jaakkola}},\ }\href {\doibase 10.1103/PhysRevLett.81.4545} {\bibfield
  {journal} {\bibinfo  {journal} {Phys. Rev. Lett.}\ }\textbf {\bibinfo
  {volume} {81}},\ \bibinfo {pages} {4545} (\bibinfo {year}
  {1998})}\BibitemShut {NoStop}%
\bibitem [{\citenamefont {Desbuquois}\ \emph {et~al.}(2012)\citenamefont
  {Desbuquois}, \citenamefont {Chomaz}, \citenamefont {Yefsah}, \citenamefont
  {L{\'e}onard}, \citenamefont {Beugnon}, \citenamefont {Weitenberg},\ and\
  \citenamefont {Dalibard}}]{Dalibard}%
  \BibitemOpen
  \bibfield  {author} {\bibinfo {author} {\bibfnamefont {R.}~\bibnamefont
  {Desbuquois}}, \bibinfo {author} {\bibfnamefont {L.}~\bibnamefont {Chomaz}},
  \bibinfo {author} {\bibfnamefont {T.}~\bibnamefont {Yefsah}}, \bibinfo
  {author} {\bibfnamefont {J.}~\bibnamefont {L{\'e}onard}}, \bibinfo {author}
  {\bibfnamefont {J.}~\bibnamefont {Beugnon}}, \bibinfo {author} {\bibfnamefont
  {C.}~\bibnamefont {Weitenberg}}, \ and\ \bibinfo {author} {\bibfnamefont
  {J.}~\bibnamefont {Dalibard}},\ }\href {\doibase 10.1038/nphys2378}
  {\bibfield  {journal} {\bibinfo  {journal} {Nature Physics}\ }\textbf
  {\bibinfo {volume} {8}},\ \bibinfo {pages} {645} (\bibinfo {year}
  {2012})}\BibitemShut {NoStop}%
\bibitem [{\citenamefont {Zhang}\ \emph {et~al.}(2008)\citenamefont {Zhang},
  \citenamefont {Tewari}, \citenamefont {Lutchyn},\ and\ \citenamefont
  {Das~Sarma}}]{SOCFermiSuperfluids}%
  \BibitemOpen
  \bibfield  {author} {\bibinfo {author} {\bibfnamefont {C.}~\bibnamefont
  {Zhang}}, \bibinfo {author} {\bibfnamefont {S.}~\bibnamefont {Tewari}},
  \bibinfo {author} {\bibfnamefont {R.~M.}\ \bibnamefont {Lutchyn}}, \ and\
  \bibinfo {author} {\bibfnamefont {S.}~\bibnamefont {Das~Sarma}},\ }\href
  {\doibase 10.1103/PhysRevLett.101.160401} {\bibfield  {journal} {\bibinfo
  {journal} {Phys. Rev. Lett.}\ }\textbf {\bibinfo {volume} {101}},\ \bibinfo
  {pages} {160401} (\bibinfo {year} {2008})}\BibitemShut {NoStop}%
\bibitem [{\citenamefont {Leggett}(1980)}]{Leggett}%
  \BibitemOpen
  \bibfield  {author} {\bibinfo {author} {\bibfnamefont {A.~J.}\ \bibnamefont
  {Leggett}},\ }in\ \href@noop {} {\emph {\bibinfo {booktitle} {Modern Trends
  in the Theory of Condensed Matter}}},\ \bibinfo {editor} {edited by\ \bibinfo
  {editor} {\bibfnamefont {A.}~\bibnamefont {P{\k{e}}kalski}}\ and\ \bibinfo
  {editor} {\bibfnamefont {J.~A.}\ \bibnamefont {Przystawa}}}\ (\bibinfo
  {publisher} {Springer Berlin Heidelberg},\ \bibinfo {address} {Berlin,
  Heidelberg},\ \bibinfo {year} {1980})\ pp.\ \bibinfo {pages}
  {13--27}\BibitemShut {NoStop}%
\bibitem [{\citenamefont {Hanai}\ \emph {et~al.}(2017)\citenamefont {Hanai},
  \citenamefont {Littlewood},\ and\ \citenamefont {Ohashi}}]{Hanai2017}%
  \BibitemOpen
  \bibfield  {author} {\bibinfo {author} {\bibfnamefont {R.}~\bibnamefont
  {Hanai}}, \bibinfo {author} {\bibfnamefont {P.~B.}\ \bibnamefont
  {Littlewood}}, \ and\ \bibinfo {author} {\bibfnamefont {Y.}~\bibnamefont
  {Ohashi}},\ }\href {\doibase 10.1103/PhysRevB.96.125206} {\bibfield
  {journal} {\bibinfo  {journal} {Phys. Rev. B}\ }\textbf {\bibinfo {volume}
  {96}},\ \bibinfo {pages} {125206} (\bibinfo {year} {2017})}\BibitemShut
  {NoStop}%
\bibitem [{\citenamefont {Samokhin}(2020)}]{Samokhin}%
  \BibitemOpen
  \bibfield  {author} {\bibinfo {author} {\bibfnamefont {K.~V.}\ \bibnamefont
  {Samokhin}},\ }\href {\doibase 10.1103/PhysRevB.101.214524} {\bibfield
  {journal} {\bibinfo  {journal} {Phys. Rev. B}\ }\textbf {\bibinfo {volume}
  {101}},\ \bibinfo {pages} {214524} (\bibinfo {year} {2020})}\BibitemShut
  {NoStop}%
\bibitem [{\citenamefont {Fil}\ and\ \citenamefont
  {Shevchenko}(2016)}]{Fil2016}%
  \BibitemOpen
  \bibfield  {author} {\bibinfo {author} {\bibfnamefont {D.~V.}\ \bibnamefont
  {Fil}}\ and\ \bibinfo {author} {\bibfnamefont {S.~I.}\ \bibnamefont
  {Shevchenko}},\ }\href {\doibase 10.1063/1.4963329} {\bibfield  {journal}
  {\bibinfo  {journal} {Low Temperature Physics}\ }\textbf {\bibinfo {volume}
  {42}},\ \bibinfo {pages} {794} (\bibinfo {year} {2016})},\ \Eprint
  {http://arxiv.org/abs/https://doi.org/10.1063/1.4963329}
  {https://doi.org/10.1063/1.4963329} \BibitemShut {NoStop}%
\end{thebibliography}%
\end{document}